\documentclass[]{article}

\usepackage[margin=0.75in]{geometry}
\usepackage[capposition=top]{floatrow}
\usepackage{xcolor}
\usepackage{amsmath}
\usepackage{amsfonts}
\usepackage{amsthm}
\usepackage{graphicx}
\usepackage{float}
\usepackage{hyperref}
\usepackage{setspace}
\usepackage{multicol}
\usepackage{lipsum}
\usepackage{algorithm}
\usepackage{algpseudocode}

\theoremstyle{definition}

\newcommand{\edit}[1]{\textcolor{black}{#1}}

\title{Unifying Statistical and Mathematical Modeling Through a Causal Inference Lens}
\author{Paul N Zivich\textsuperscript{1}}
\date{
	\small
	\textsuperscript{1}Department of Epidemiology, Gillings School of Global Public Health, University of North Carolina, Chapel Hill, NC\\
	~\\
	\today
}

\begin{document}
	
\maketitle

\begin{abstract}
	Within the biological, physical, and social sciences, there are two broad quantitative traditions: statistical and mathematical modeling. Both traditions have the common pursuit of advancing our scientific knowledge, but these traditions have developed largely with distinct languages and inferential frameworks. This paper uses the notion of identification from causal inference, a field originating from the statistical modeling tradition, to develop a shared language. I first review foundational identification results for statistical models and then extend these ideas to mathematical models. Central to this framework is the use of bounds, ranges of plausible numerical values, to analyze both statistical and mathematical models. I discuss the implications of this perspective for the interpretation, comparison, and integration of different modeling approaches, and illustrate the framework with a simple pharmacodynamic model for hypertension. To conclude, I describe areas where the approach taken here should be extended in the future. By formalizing connections between statistical and mathematical modeling, this work contributes to a shared framework for quantitative science. My hope is that this work will  advance interactions between these two traditions.
\end{abstract}

\section*{Introduction}

There are broadly two quantitative traditions in the biological, physical, and social sciences \cite{mayUsesAbusesMathematics2004, kuhneCausalEvidenceHealth2022, pengComparisonPhenomenologicalApproach2022, naimiCommentaryIntegratingComplex2016a, transtrumBridgingMechanisticPhenomenological2016, weinbergerMakingSenseNonfactual2020, whiteShouldWeCare2019, christopheComputationalModelsNeurosciences2022, bakerMechanisticModelsMachine2018}. The first \edit{has been often referred} to as statistical, empirical, or phenomenological modeling. Statistical modeling relies on using empirical observations to 
\edit{solve for parameters that produce optimal predictions, where optimal defined by a chosen loss-function. Examples of the statistical modeling tradition include} randomized trials, regression, structural equation modeling, and g-methods \cite{naimiIntroductionMethods2017, robinsNewApproachCausal1986}. The second, I will refer to as mathematical or mechanistic modeling. Mathematical modeling instead proceeds from a \edit{hypothesized mechanistic} representation of a system through formulas that does not necessitate direct \edit{empirical} observations. Examples of this approach include dynamic (e.g., differential equations), compartmental, microsimulation, and agent-based models \cite{bjornstadModelingInfectiousEpidemics2020, bjornstadSEIRSModelInfectious2020, krijkampMicrosimulationModelingHealth2018, caglayanMicrosimulationModelingOncology2018, el-sayedSocialNetworkAnalysis2012b, tedeschiReviewPrevailingMathematical2023}. 
\edit{To highlight the differences between these traditions, a few illustrative examples are presented. First, is modeling the period of a pendulum relative to its length, which can be done through derivation of equations from Newton's Laws (mathematical modeling) or through direct observations (statistical modeling) \cite{tolasaTheoreticalAnalysisSimple2025}. A second example comes from weather forecasting. Traditionally, physics-based models have been used to simulate future weather patterns (mathematical modeling) \cite{bauerQuietRevolutionNumerical2015}. More recently, there has been interest in the use of machine learning trained on historical data (statistical modeling) \cite{zhangMachineLearningMethods2025}. These traditions have also been contrasted in the context of infectious disease forecasting \cite{heesterbeekModelingInfectiousDisease2015, miyamaPhenomenologicalMechanisticModels2022}. A third example comes from stem cells, where there has been intellectual conflict between theoretical biologists (mathematical modeling) and experimentalists (statistical modeling) \cite{faganStemCellsSystems2016}. A fourth example is in approaches to studying how hand hygiene interventions can mitigate disease spread in office environments. One approach has been to compare outcomes between offices with and without different interventions (statistical modeling) \cite{arbogastImpactComprehensiveWorkplace2016a}. Alternatively, others have used modeled pathogen-to-fomite-to-person transmissions to project efficacy of increased hand hygiene (mathematical modeling) \cite{beamerModelingHumanViruses2015a}. A final example comes from medical research, which has traditionally relied on randomized trials to study the efficacy therapies but there is also interest in simulated trials based on pharmacokinetics (mathematical modeling) \cite{holfordClinicalTrialSimulation2010b}. While there is no sharp division between these traditions \cite{whiteShouldWeCare2019, christopheComputationalModelsNeurosciences2022, ackleyDynamicalModelingTool2022}, viewing these as separate traditions clarifies the complementary roles each plays in contributing} to scientific knowledge. \edit{As with competing methods more generally \cite{breimanStatisticalModelingTwo2001a, kennedyCommentStatisticalModeling2021}, drawing a distinction can help to understand when one approach may be preferable to or more applicable than the other \cite{bakerMechanisticModelsMachine2018, whiteShouldWeCare2019, ackleyDynamicalModelingTool2022}.}

Given the broadness of this topic, I use some of the recent epidemiologic literature as an entry point \edit{into deeper discussion of these comparisons}. Murray et al. provided a comparison between a particular statistical model developed for causal inference, the parametric g-formula, to \edit{a mathematical model,} microsimulation, for inferring time-varying causal effects \cite{murrayComparisonAgentBasedModels2017}. \edit{The authors conclude that mathematical models face challenges to parameterization in settings with multiple time-varying variables.} That research inspired a wider discussion \edit{of the role different modeling strategies play} in epidemiology \cite{edwardsInvitedCommentaryCausal2017, keyesInvitedCommentaryAgentBased2017c}. \edit{Subsequent work has further contrasted results from other statistical and mathematical models in specific applications \cite{mooneyGComputationAgentBasedModeling2022}, discussed the similarities between graphical causal models mechanistic models \cite{arnoldDAGinformedRegressionModelling2018, mooijOrdinaryDifferentialEquations2013, aalenCanWeBelieve2016, wordenProductsCompartmentalModels2017, havumakiUsingCompartmentalModels2020}, and described how each approach can be used to study the effect of interventions \cite{ackleyDynamicalModelingTool2022}.} However, less work has focused on studying both traditions from a shared perspective, with some exceptions. \edit{First, Marshall and Galea describe how an agent-based model can be used to estimate causal effects in public health settings by making an analogy to the correct model specification assumption used in statistical modeling \cite{marshallFormalizingRoleAgentBased2015}. In my own work, co-authors and I have proposed the use of mathematical models to address data gaps that statistical models face \cite{zivichSynthesisEstimatorsTransportability2025, zivichSynthesisEstimatorsTransportability2025}. Implicitly, this relies on mathematical models identifying causal effects, which was done through the assumption that the mathematical model captures the relevant structure and parameter variation \cite{zivichSynthesisEstimatorsTransportability2025}. More thorough formulations of these ideas may help to clarify some of the existing ambiguities around the assumptions being made, in particular when contrasting the assumptions used by each modeling strategy.}

Here, I consider how mathematical models \edit{can be analyzed} using tools from causal inference, a field growing out of the statistical modeling tradition that is popular in the medical and social sciences \cite{dahabrehCausalInferenceEffects2024, imbensCausalInferenceSocial2024}. 
\edit{While statistical and mathematical models can be used for other purposes \cite{itoDistinguishingDescriptionPrediction2025, wintherMathematicalModelingBiology2012}, I focus here on the use of time-dependent models for determining the effect of an action on some outcome.} The proposed framework allows more direct comparisons between statistical and mathematical models through a shared language \edit{for causal effects}, which has been previously called for in epidemiology \cite{ackleyDynamicalModelingTool2022}. My approach to linking the two methodological traditions is through bounds \cite{manskiNonparametricBoundsTreatment1990}, or ranges of plausible numeric values, that address a \edit{pre-specified} scientific question. First, I review some known results regarding the bound for statistical models for causal effects. Next, I consider how to construct bounds for mathematical models. Some implications of these bounds are reviewed, and then I apply this framework with a simple case study to illustrate the potential of these ideas.

\section*{Interest Parameter}

To help fix ideas, suppose our motivating scientific equation was to learn the effect of the tetanus vaccine on symptomatic tetanus in a specific population. One of the first steps in quantitative research is to translate this general scientific question into one that can be addressed through a numerical value, often referred to as the interest parameter \cite{dangCausalRoadmapGenerating2023}. An interest parameter that addresses this motivating question is the the average causal effect (ACE) \cite{hudgensCausalInferenceInterference2008a, savjeAverageTreatmentEffects2021}, which contrasts the case where everyone received the vaccine versus no one received the vaccine.

Here, capital letters will be used to denote random variables with lowercase denoting possible realizations. Calligraphic letters denote the support, or the set of possible values that a random variable could take. Lowercase Greek letters will denote parameters with uppercase Greek letters denoting the parameter space (i.e., the set of possible values the parameter can take). To distinguish between equality and assignment, $:=$ will be used to denote assignment. To define causal effects, I will use Neyman potential outcomes \cite{splawa-neymanApplicationProbabilityTheory1990}, where $Y_i^a$ denotes the outcome unit $i$ will have under vaccination status $a$, with $a$ denoting tetanus vaccination and $0$ denoting no vaccination for unit $i$. Therefore, each person has two potential outcomes in this context. Note that potential outcomes cannot be directly observed and can be thought of as a conceptual device with which causal effects can be defined. As tetanus is not considered to be human-to-human transmissible \cite{palAnimalHumanTetanus2024}, we can reasonably assume a person's potential outcome only depends on their vaccination status (i.e., no interference). The ACE can then be written as $\psi := \mu_1 - \mu_0$, where $\mu_a := \Pr_{S=1}(Y^a = 1)$ and $\Pr_{S=1}(\cdot)$ is the probability function. \edit{Here, the} $S=1$ \edit{subscript is used as a reminder that the probability is only} defined for a particular context or population located in time and space. Normally \edit{this context} is left implicit, but here it is explicitly written for \edit{emphasis}.

Note that $\psi$ must take a value within the range $[-1,1]$ following the axioms of probability \cite{kolmogorovFoundationsTheoryProbability1950}. This set of possible values for $\psi$ (i.e., the parameter space) is denoted by $\Psi$. As potential outcomes are not directly observed, we consider assumptions that enable $\psi$ to be learned.

\section*{Identification}

\begin{figure}
	\centering
	\caption{Visualization of different types of identification results}
	\includegraphics[width=0.7\linewidth]{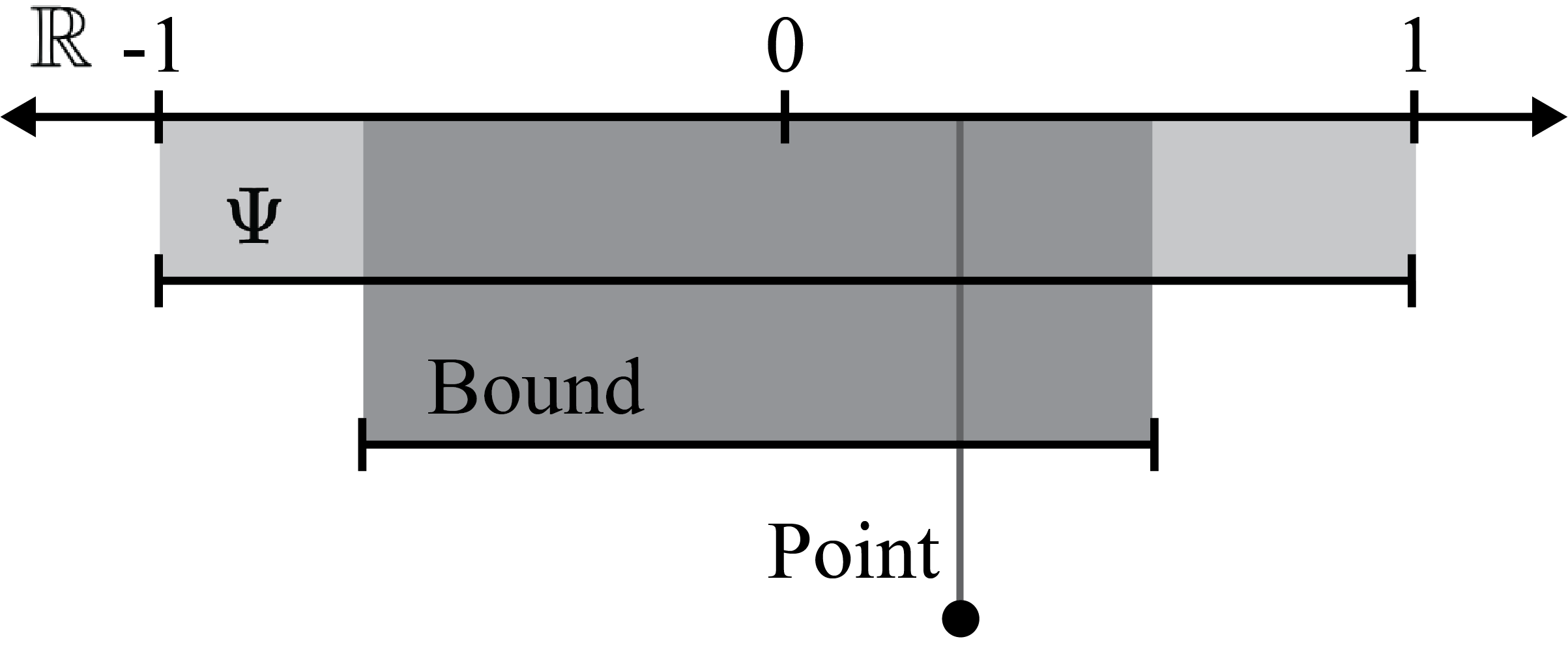}
	\floatfoot{$\Psi$ is the parameter space, which for the average causal effect of a binary outcome is $[-1,1]$.}
	\label{Fig1}
\end{figure}

Within the field of causal inference, the next task is to discern under what conditions the interest parameter would be identified, or expressed as a function of the observed data \cite{dangCausalRoadmapGenerating2023, aronowNonparametricIdentificationNot2025}. The steps that license this re-expression of our causal parameter in terms of the observable data can then be understood as claims about the world that allow for a causal interpretation of a quantitative analysis \cite{robinsAssociationCausationMarginal1999}. Identification can be further delineated. First, identification can be \textit{nonparametric} or \textit{parametric}. This modifier refers to whether there are parametric constraints placed on \edit{relationships between} the observed \edit{variables}. A second classification is \textit{partial} versus \textit{point} identification. Partial identification indicates that at best one can only determine a range of values for $\psi$ that is proper subset of $\Psi$. This range of values is commonly referred to as the bounds (Figure \ref{Fig1}) \cite{manskiNonparametricBoundsTreatment1990}. Alternatively, point identification provides a unique mapping of $\psi$ to a single value.

For my analysis of mathematical models, the prior definition of identification needs to be expanded. There is prior work on the identifiability of parameters in mathematical models, like SIR models, with observed data \cite{eisenbergDeterminingIdentifiableParameter2014, kabanikhinIdentifiabilityMathematicalModels2016, krivorotkoSensitivityAnalysisPractical2020, kabanikhinPracticalIdentifiabilityMathematical2021, carpioParameterIdentificationEpidemiological2022}. Given these models have parameters estimated using observations, this prior work might be better seen as being part of the statistical modeling tradition (at least from the perspective I take here). As stated, my conceptualization of a mathematical model does not require observations. To address this deficiency in the definition of identification, I instead define identification as the task of expressing the interest parameter as a function of the available \textit{information}, where observations or theory are simply types of information. 

\subsection*{Statistical Models}

A few important identification results for $\psi$ from statistical modeling are reviewed. Let $Y_i$ denote symptomatic tetanus (1: yes, 0: no) and $A_i$ indicate vaccination (1: yes, 0: no) for person $i$. Suppose we observed independent and identically distributed observations $O_i = (A_i, Y_i)$ for a random sample of the population $S=1$. Further, suppose that $A_i$ precedes $Y_i$ in time (which can be guaranteed by design) and $O_i$ is measured without error. Here, $\psi$ is identified if we can write it as a function of $O_i$. Given observations are independent, the index subscript is left implicit hereafter. Note that 
\begin{equation*}
	\begin{aligned}
		\mu_a = & \Pr_{S=1}(Y^a = 1 \mid A=a) \Pr_{S=1}(A=a) + \Pr_{S=1}(Y^a = 1 \mid A \ne a) \Pr_{S=1}(A \ne a) \\
		= & \Pr_{S=1}(Y = 1 \mid A=a) \Pr_{S=1}(A=a) + \Pr_{S=1}(Y^a = 1 \mid A \ne a) \Pr_{S=1}(A \ne a)
	\end{aligned}
\end{equation*}
follows from the the law of total probability and causal consistency (i.e., $Y^{A_i}_i = Y_i$) \cite{coleConsistencyStatementCausal2009, vanderweeleConcerningConsistencyAssumption2009}, respectively. Causal consistency can be thought of as a primitive connection between the hypothetical potential outcome $Y_i^a$ and the observed outcome $Y_i$. Namely, this assumption asserts that the observed outcome is the potential outcome under the action that occurred.

All but one part of the previous expression is written in terms of observable data, namely $\Pr_{S=1}(Y^a = 1 \mid A \ne a)$. However, we know from probability theory that this quantity must be within $[0,1]$. Therefore, $\Pr_{S=1}(Y^a = 1 \mid A \ne a)$ can be set to \edit{the boundary points} to obtain the following nonparametric partial identification result
\begin{equation*}
	\mu_a \in \left[ \Pr_{S=1}(Y=1 \mid A=a) \Pr_{S=1}(A=a), \;\; 
	                 \Pr_{S=1}(Y=1 \mid A=a) \Pr_{S=1}(A=a) + \Pr_{S=1}(A \ne a) \right]
\end{equation*}
Repeating this process for $a \in \{0, 1\}$, the nonparametric bounds for $\psi$ are then
\begin{equation*}
	\psi \in \left[-\Pr_{S=1}(Y=0, A=1) - \Pr_{S=1}(Y=1, A=0), \;\;
		           \Pr_{S=1}(Y=1, A=1) - \Pr_{S=1}(Y=0, A=0) 
	\right].
\end{equation*}
Note that these bounds always have a width of 1, which is a halving of the original parameter space. However, these unit-length bounds must contain the null of no effect of $A$ on $Y$. To narrow the bounds further, additional assumptions are needed.

Consider the additional assumption that the observations originated from a trial that assigned values of $A$ based on unconditional randomization. Randomization implies marginal exchangeability (i.e., $Y^a$ is marginally independent of $A$) with positivity (i.e., there is a non-zero chance of receiving either $A=1$ or $A=0$) is given by design \cite{hernanEstimatingCausalEffects2006}. Together these assumptions imply that 
\begin{equation*}
	\Pr_{S=1}(Y^a = 1 \mid A=a) = \Pr_{S=1}(Y^a = 1) = \Pr_{S=1}(Y^a = 1 \mid A \ne a)
\end{equation*}
for $a \in \{0,1\}$. Therefore, $\mu_a$ is equal to $\Pr_{S=1}(Y = 1 \mid A=a)$ and $\psi$ is nonparametrically point-identified as $\Pr_{S=1}(Y = 1 \mid A=1) - \Pr_{S=1}(Y = 1 \mid A=0)$. In the absence of randomization, other approaches to deal with $\Pr_{S=1}(Y^a = 1 \mid A \ne a)$ are needed for point identification of $\psi$. This has been the focus of research on causal inference with observational data \cite{hernanEstimatingCausalEffects2006, tchetgenIntroductionProximalCausal2024a, zivichINTRODUCINGPROXIMALCAUSAL2023}. One common assumption to progress is the assumption of no uncontrolled confounding given a set of measure variables \cite{hernanEstimatingCausalEffects2006}. If we let $W_i$ denote this set of variables, it follows from conditional exchangeability by $W_i$ (i.e., no uncontrolled confounding given $W_i$)  with positivity that 
\begin{equation*}
	\mu_a = \int_{\mathcal{W}} \Pr_{S=1}(Y=1 \mid A=a, W=w) f_{W|S=1}(w)
\end{equation*}
where $f_{W|S=1}(\cdot)$ is the probability density function for $W$ in $S=1$. This identification result is commonly referred to as the g-formula \cite{robinsNewApproachCausal1986}. These identification results can then be used to develop an estimator based on a statistical model. Here, we might consider a logistic regression model for $\Pr_{S=1}(Y = 1 \mid A=a,W=w)$ denoted by $m(A,W;\gamma)$ where $\gamma$ is the parameter vector for the statistical model. The parametric g-formula estimator is then
\begin{equation*}
	\hat{\mu}_a := \sum_{i=1}^{n} m(A_i,W_i; \hat{\gamma})
\end{equation*}
where \edit{$i$ indexes the individual units of observation and} hats denote parameters estimated using the observed data. To replace the probability function with $m(A,W;\gamma)$, an additional assumption commonly referred to as correct (statistical) model specification is relied on \cite{aronowNonparametricIdentificationNot2025, vansteelandtModelSelectionModel2012, maclarenWhatCanBe2020}. This assumption stipulates that the true data-generating probability function for $Y$ given $A,W$ is contained within the set of probability distributions allowed by the statistical model, i.e., $\Pr_{S=1}(Y=1 \mid A=a, W=w) \in \{m(a,w;\gamma) : \gamma \in \Gamma\}$.

\subsubsection*{Implications}

This framework offers some important insights on how statistical models might be evaluated by articulating their assumptions that license a causal interpretation. Interestingly, simply observing a random sample of $A,Y$ under the assumptions of causal consistency and no measurement error allows us to narrow the set of possible values for $\psi$. As noted though, these possible values will always include `no effect'. As a result, auxiliary assumptions to further narrow the bounds are important to reach more definitive conclusions. From this perspective, randomized trials are special in that they satisfy additional assumptions by design. Absent this design-based feature, we must rely on unverifiable assumptions instead. Beyond the identification assumptions, use of a statistical model brings along a correct statistical model specification assumption for estimation. If a saturated model is used (i.e., the model has as many parameters as unique combinations of values for $A,W$), then the correct model specification assumption automatically holds. However, fitting saturated models in practice is often infeasible \edit{due to the curse of dimensionality}, so correct model specification is often relied on. This additional assumption for estimation beyond those of the identification assumptions have been previously discussed as the gap between identification and (statistical) estimation \cite{maclarenWhatCanBe2020}, and used to highlight why randomized trials are special from a design perspective \cite{aronowNonparametricIdentificationNot2025}. Not reviewed here, but there are re-expressions and extensions of the g-formula which rely on different statistical modeling assumptions \cite{tchetgenIntroductionProximalCausal2024a, robinsAssociationCausationMarginal1999}. Importantly, some of these modifications allow for weakened versions of the correct model specification assumption, whereby multiple distinct models are specified and only one needs to be correctly specified for estimation \cite{bangDoublyRobustEstimation2005a}. Much of the recent interest in machine learning for causal effects is driven by interest in making the correct model specification assumption more plausible \cite{zivichMachineLearningCausal2022}. Finally, the notion of identification can also be used to clarify assumptions for methods addressing other systematic errors, like selection bias or missing data \cite{westreichBerksonsBiasSelection2012, edwardsAllYourData2015}.

\subsection*{Mathematical Models}

\begin{figure}
	\centering
	\caption{Directed acyclic graph representing the vaccine, immune, infection mechanism}
	\includegraphics[width=0.5\linewidth]{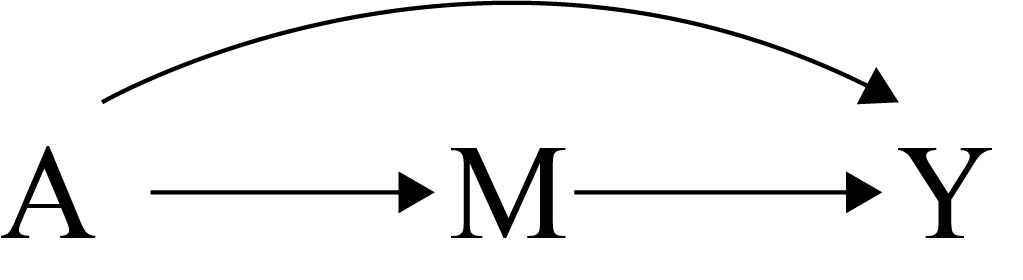}
	\floatfoot{$A$: vaccination, $M$: immune system response, $Y$: infection}
	\label{Fig2}
\end{figure}

Now consider the following process for developing a mathematical model for the effect of vaccination on symptomatic tetanus infection. 
\edit{While there is extensive guidance on constructing mathematical models \cite{krijkampMicrosimulationModelingHealth2018, robertsConceptualizingModelReport2012, robinsonTenRulesEffective2022}, I set the particulars of that guidance aside for the following discussion.} First, \edit{I} specify a mechanism between $A$ (vaccination) and $Y$ (symptomatic infection). For expository purposes, a relatively simple mechanism is considered. Here, a single variable $M$, which denotes an immune system response to vaccination (1: response, 0: none) is considered. Like the statistical modeling above, the set of variables is assumed to be well-defined for $S=1$ and have a clear time-ordering (i.e., $A \rightarrow M \rightarrow Y$). Following this time-ordering, a complete (i.e., all possible edges) directed acyclic graph can be used to represent this causal mechanism (Figure \ref{Fig2}).

Given this structure, $\mu_a$ can be decomposed as
\begin{equation*}
	\mu_a = \sum_{m \in \{0, 1\}} \Pr_{S=1}(Y^a = 1 \mid M^a = m) \Pr_{S=1}(M^a = m)
\end{equation*}
where $\Pr_{S=1}(A=a) := 1$ for the causal mean when $A$ is set to $a$ and zero otherwise. Therefore, this mechanistic model is composed of two functions: a function for antibody response given vaccination and a function for symptomatic tetanus infection status given antibody response and vaccination. As with a statistical model, some function for the processes must be chosen. Let $g(a; \theta)$ be a function chosen to generate antibody response defined by the $j$-dimension parameter vector $\theta$. Let $h(a,m; \lambda)$ be a function chosen to generate infection status defined by the $k$-dimension parameter vector $\lambda$. Some restrictions are placed on the chosen functions. First, the function $g$ returns values between zero and one (i.e., $g: \Theta \rightarrow [0,1]$). Similarly, we have $h: \Lambda \rightarrow [0,1]$. Given these functions, $\psi$ can be computed via
\begin{equation*}
	\bar{\mu}_a(\theta, \lambda) := \sum_{m \in \{0, 1\}} h(a,m;\lambda)g(a; \theta)
\end{equation*}
where the overbar is to emphasize the distinction between the computational procedure for the mathematical model and the parameter the mathematical model is meant to compute, $\mu_a$. The computed ACE from the mathematical model is then $\bar{\psi}(\theta, \lambda) := \bar{\mu}_1(\theta, \lambda) - \bar{\mu}_0(\theta, \lambda)$.

\begin{figure}
	\centering
	\caption{Visualization of the parameter space covered by a vacuous versus non-vacuous model}
	\includegraphics[width=0.6\linewidth]{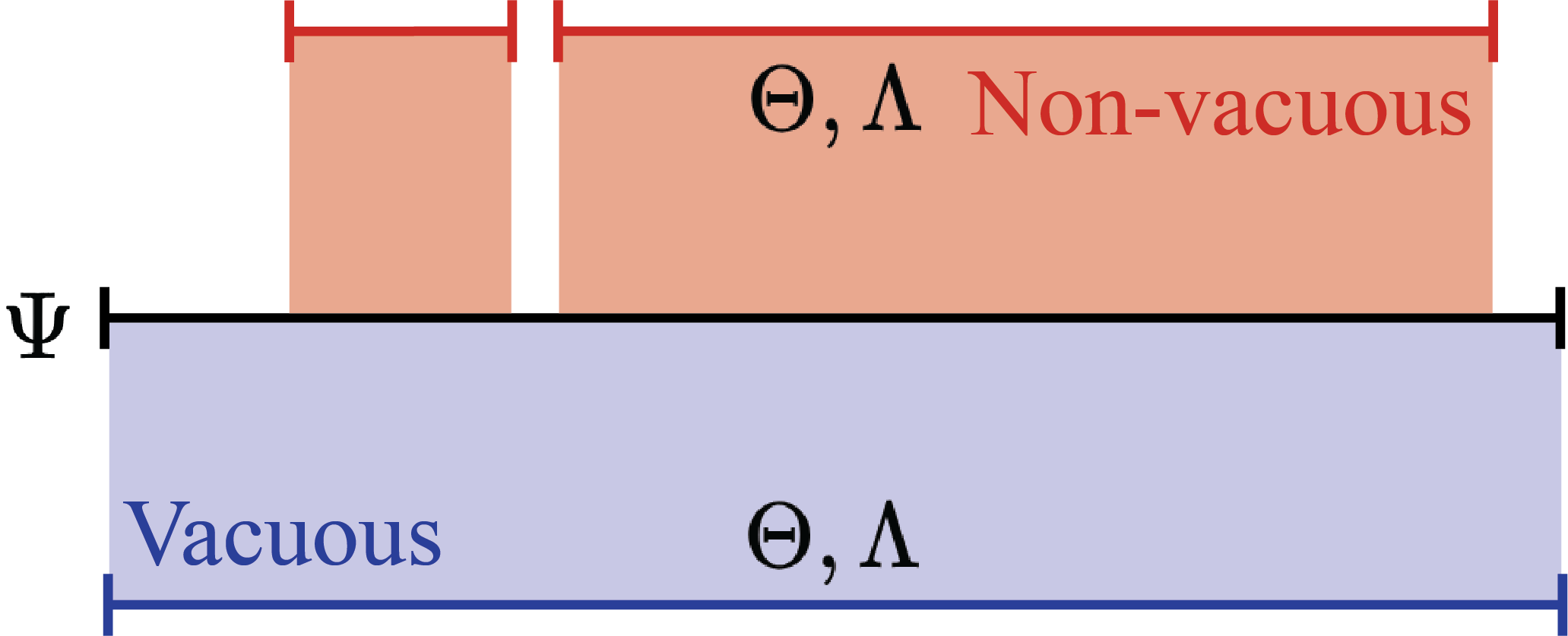}
	\floatfoot{$\Psi$ denotes the parameter space. $\Theta,\Lambda$ denote the parameter space for the chosen functions in the corresponding mathematical model.}
	\label{Fig3}
\end{figure}

As indicated by this model, specific parameter values for $\theta$ and $\lambda$ must be chosen to compute $\psi$. Before considering how $\bar{\psi}(\theta, \lambda)$ changes for the input $\theta \in \Theta$ and $\lambda \in \Lambda$, some additional constraints are placed on the function $g$ and $h$. Here, the output of $g$ (i.e., image) covers the interval from zero to one for all unique values of $a$ (i.e., $\{g(a; \theta) : \theta \in \Theta\} = [0,1]$ for $a \in \mathcal{A}$). Similarly, $\{h(a,m; \lambda) : \lambda \in \Lambda\} = [0,1]$ for $a,m \in \mathcal{A} \times \mathcal{B}$. As examples of functions that satisfy these properties, consider $g(a;\theta) := \text{expit}(\theta_0 + \theta_1 a)$ with $\theta := (\theta_0, \theta_1) \in \mathbb{R}^2$ and $h(a,m;\lambda) := \text{expit}(\lambda_0 + \lambda_1 a + \lambda_2 m + \lambda_3 a m)$ with $\lambda := (\lambda_0, \lambda_1, \lambda_2, \lambda_3) \in \mathbb{R}^4$, where $\text{expit}(x) := \{1 + \exp(-x)\}^{-1}$. When adding the boundary points of zero and one, the set of outputs from the mathematical model will span the parameter space, i.e., $\{\bar{\psi}(\theta, \lambda) : \theta, \lambda \in \Theta \times \Lambda\} \cup \{-1, 0, 1\} = [-1,1] = \Psi$, for any input $a,m$ combination. I refer to a mathematical model that satisfies this property as \textit{vacuous}, since such a model provides no information about $\psi$ beyond that it lies within the parameter space for any value of $a$ and $m$ (which was known prior to constructing the model). Models that do not satisfy this property are referred to as \textit{non-vacuous}. A vacuous model separates the implications of function choices from the parameter choices. Structural assumptions about the mechanism are thus encoded in the choice of parameters (e.g., $A$ not having a direct effect on $Y$ would be encoded by $\text{expit}(\lambda_0 + 0 a + \lambda_2 m + 0 am)$) and not by the choice of functions. Non-vacuous mathematical models instead rule out certain values for $\psi$ before even considering values of $\theta$ and $\lambda$ (Figure \ref{Fig3}). \edit{As an example of a non-vacuous model, consider the SARS-CoV-2 vaccination model of Padmanabhan et al. \cite{padmanabhanModelingHowAntibody2022}. The chosen mathematical model structure restricts the mechanism of vaccination through an antibody response (i.e., the arrow from $A$ to $Y$ in Figure \ref{Fig2} is not present). For this model to become vacuous, the model structure would need to allow for vaccination to affect the outcome through pathways outside of the antibody response (e.g., cell-mediated immunity).}

To progress, suppose that an oracle revealed the true values of $\theta$ and $\lambda$ for our population, denoted by $\theta^*$ and $\lambda^*$. Therefore, we know that
\begin{equation}
	\begin{aligned}
		\Pr_{S=1}(M^a = m) & = g(a; \theta^*) \\
		\Pr_{S=1}(Y^a = 1 \mid M^a = m) & = h(a,m; \lambda^*) \\
	\end{aligned}
	\label{Eq1}
\end{equation}
which means that $\psi = \bar{\psi}(\theta^*, \lambda^*)$. This result is a point identification result for the mathematical model. Note that the vacuous model means that $\theta^*$ and $\lambda^*$ that satisfy this equality exist. In the case of a non-vacuous model, it may not be possible for \eqref{Eq1} to hold for any $\theta,\lambda \in \Theta \times \Lambda$.

\begin{figure}
	\centering
	\caption{Visualization of the proposed types of identification results for mathematical models}
	\includegraphics[width=0.7\linewidth]{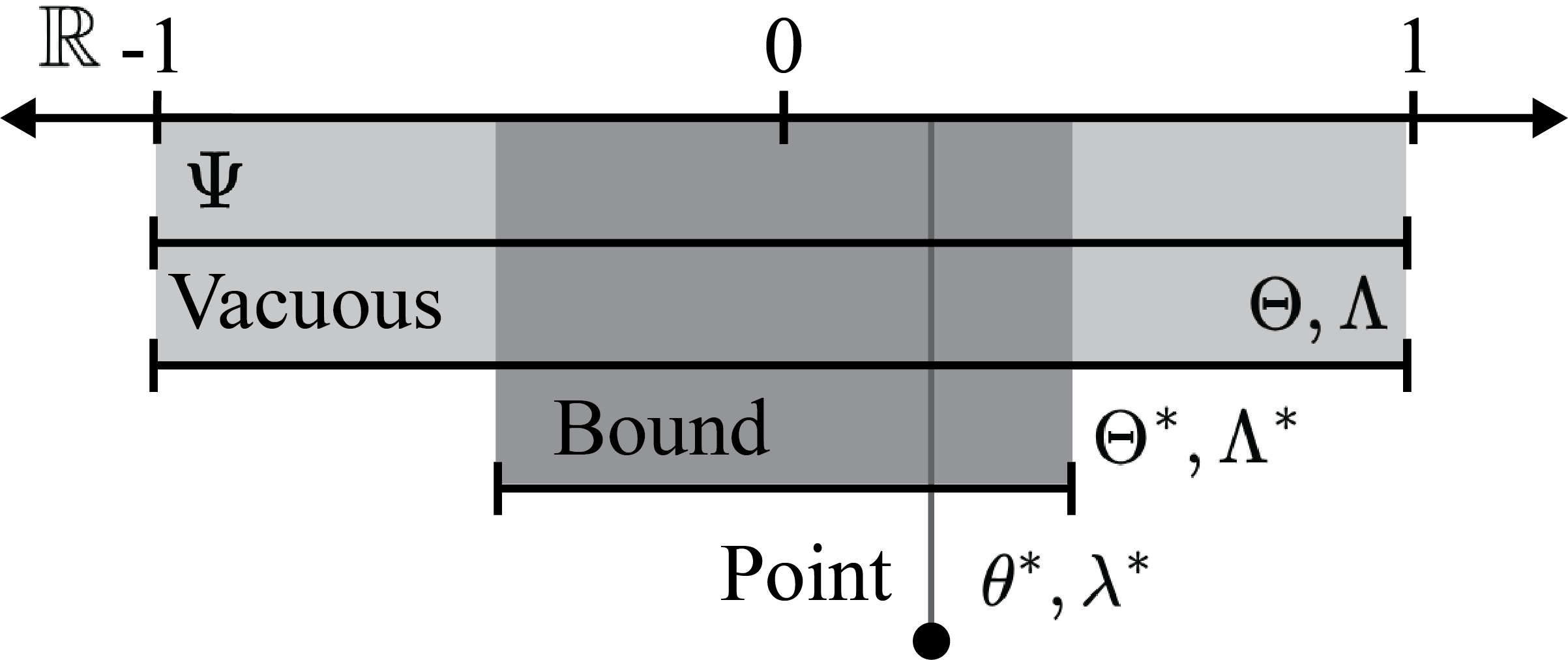}
	\floatfoot{$\Psi$ is the parameter space, which for the average causal effect of a binary outcome is $[-1,1]$. $\Theta,\Lambda$ denote the parameter space for the chosen functions in the corresponding mathematical model. $\Theta^*,\Lambda^*$ denote the selected subset of parameter values. $\theta^*,\lambda^*$ denote the true parameter values.}
	\label{Fig4}
\end{figure}

In absence of this mythical oracle, one may still partially identify $\psi$. Let $\Theta^*$ be a range of values that is a strict subset of $\Theta$. Similarly let $\Lambda^*$ be a range of values that is a strict subset of $\Lambda$. Then assume that 
\begin{equation}
	\begin{aligned}
		\Pr_{S=1}(M^a = m) & \in \{g(a; \theta) : \theta \in \Theta^*\} \\
		\Pr_{S=1}(Y^a = 1 \mid M^a = m) & \in \{h(a,m; \lambda) : \lambda \in \Lambda^*\} \\
	\end{aligned}
	\label{Eq2}
\end{equation}
or equivalently that $\theta^* \in \Theta^*$ and $\lambda^* \in \Lambda^*$. Since the phrase `correct model specification' is already in use for statistical modeling, I propose to refer to \eqref{Eq2} as the `model-capture' assumption, since the functions are meant to capture the true probabilities. Also note that \eqref{Eq1} is a special case of \eqref{Eq2}, where the sets in \eqref{Eq2} each only contain a single value. Under the model-capture assumption, it follows that
\begin{equation*}
	\psi \in \left[ \min_{\theta,\lambda \in \Theta^* \times \Lambda^*} \bar{\psi}(\theta, \lambda), 
	\;  \;
	\max_{\theta,\lambda \in \Theta^* \times \Lambda^*} \bar{\psi}(\theta, \lambda) \right] = \Psi^* \subset \Psi
\end{equation*}
which is a partial identification result for $\psi$. For intuition behind these varying results, see Figure \ref{Fig4}

\subsubsection*{Implications}

The previous framework for analyzing a mathematical model offers some important insights on how mathematical models might be evaluated regarding their utility. Here, the utility of a mathematical model for constructing bounds for a causal effect in a given context depends on the model-capture assumption in \eqref{Eq2}. Importantly, the role between functions and parameters was separated by introducing the concept of a \textit{vacuous model} which meant that no structural assumptions were imposed beyond the definition of variables and time-ordering. There is an analog to problem faced by non-vacuous models in statistical modeling: the g-null paradox, where the parametric g-formula is guaranteed to be misspecified under the sharp causal null hypothesis with time-varying confounding \cite{mcgrathRevisitingGnullParadox2022}. The g-null paradox is an undesirable property of the parametric g-formula. I suggest that non-vacuous mathematical models should be viewed in a similar manner to the g-null paradox, undesirable but acceptable in absence of alternatives.

Equation \eqref{Eq2} also indicates how a mathematical model might be justified or evaluated. Given a context, choices of $\Theta^*$ and $\Lambda^*$ need to be justified. Here, external information (e.g., trials, observational cohorts, pharmacokinetic studies, animal models, beliefs, etc.) that is relevant to the context $S=1$ should guide these choices. Specifying narrow ranges for $\Theta^*,\Lambda^*$ (and thus computing narrow bounds for $\psi$) is only justifiable given the available information. Like the no unmeasured confounding assumption in statistical modeling with observational data, this assumption is not given by study design. To illustrate, consider the mathematical model developed to evaluate the effect of a hand washing intervention on rotavirus and rhinovirus transmission among adults \cite{beamerModelingHumanViruses2015a}. This model relied on a parameter for hand-to-mouth contacts, with the choice based on data from children aged 7-12 years old. As noted by the authors, this choice might be a conservative value for this parameter (but was unverified) and thus the results should be viewed with that in mind. The framework presented here helps to formalize these notions. \edit{Additionally, this framework might be helpful to clarify common adages in modeling , like simpler models are to be preferred \cite{robinsonTenRulesEffective2022, coenHowMathematicalModels2007}.} 

Finally, this framework may help to resolve some ongoing debates in epidemiology about the exchangeability assumption. Prior literature has argued whether exchangeability plays a role for mathematical models \cite{naimiCommentaryIntegratingComplex2016a}, with several authors claiming that exchangeability does not \cite{el-sayedSocialNetworkAnalysis2012b, marshallFormalizingRoleAgentBased2015, lofgrenReIntegratingComplex2017}. As a modeler directly sets $\Pr_{S=1}(A=a)$ exchangeability for $A$ is met by design, much like a randomized trial, because it is under direct control of the investigator. However, exchangeability does still have a role to play with mathematical models. From the perspective espoused here, the external information used to justify the model-capture assumption needs to be relevant, or generalizable, to the context $S=1$. 
In the absence of this marginal exchangeability of background information between contexts, the information used to narrow the possible parameter values would no longer justify \eqref{Eq2}. \edit{For example, if the external information for vaccination on immune response came from adults but we want the model to estimate the efficacy in children, then we might be skeptical of the model-capture assumption.} The movement from the output of a computational model to making decisions in a human population described by Naimi is what is captured by this exchangeability assumption \cite{naimiCommentaryIntegratingComplex2016a}. 

\section*{Case Study}

To contextualize the previous ideas, I develop a simple mathematical model for amlodipine, a calcium channel blocker used to treat high blood pressure and chest pain, on systolic blood pressure. Note that this model is only intended to frame the prior concepts. Here, the motivating scientific question is ``what is the difference in the risk of hypertension (defined as a systolic blood pressure of $\ge140$ mm Hg) 24 hours after taking amlodipine (10 mg) or placebo among adults 18 years or older with a systolic blood pressure above 140 in the United States?". This parameter is again denoted by $\psi$, where $Y^a$ now denotes the potential hypertension under treatment $a$ (1: 10 mg amlodipine, 0: placebo) and the context $S=1$ is adults 18 years or older with a systolic blood pressure above 140 in the United States. 

\begin{figure}
	\centering
	\caption{Assumed mechanism between amlodipine and subsequent systolic blood pressure at 24 hours}
	\includegraphics[width=0.5\linewidth]{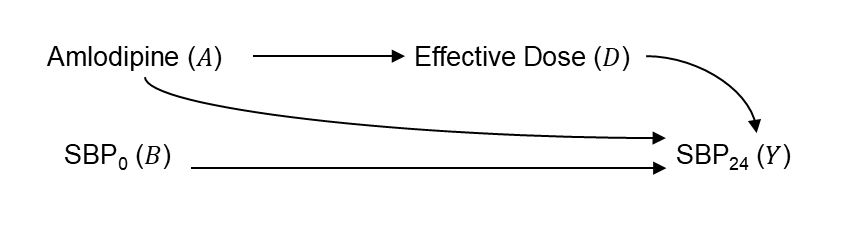}
	\floatfoot{SBP: systolic blood pressure, with subscripts indicating time since baseline.}
	\label{Fig5}
\end{figure}

To identify and estimate $\psi$, a relative simple mathematical model is considered. Here, the following pharmacodynamic model based on the structure in Figure \ref{Fig5} is used
\begin{equation}
	\bar{\mu}_a := \int_{\mathcal{B}} h(b,a,g(\theta); \lambda) f(b)
	\label{Eq3}
\end{equation}
where $a$ denotes the assigned treatment and $b$ denotes the baseline systolic blood pressure with the support $\mathcal{B}$. This model is composed of three functions: $f$ is the probability density function for baseline systolic blood pressure of the population, $g$ is a function for the effective concentration of amlodipine at 24 hours, and $h$ is a function for systolic blood pressure at 24 hours. Here, $f$ is left as a generic probability density function that satisfies the constraint $f(w) = 0$ for $w < 140$. To model the effective concentration, a simple proportion model is used
\begin{equation*}
	g(\theta) := \theta_0 \times \theta_1
\end{equation*}
where $\theta_0$ is the dose (in mg) and $\theta_1$ is the percentage of active drug remaining at 24 hours. For hypertension, the following model based on a Hill Equation (E-max) model is used
\begin{equation*}
	h(b,a,m; \lambda) := I(\bar{y}(b, a, m; \lambda) < 140)
\end{equation*}
where 
\begin{equation*}
	\bar{y}(b,a,m; \lambda) := b - \left[ \lambda_0 + a \lambda_1 \left(\frac{m}{\lambda_2 + m}\right) + a \lambda_3 \right],
\end{equation*}
$I(\cdot)$ is the indicator function, $\lambda_0$ is a constant change in the baseline systolic blood pressure, $\lambda_1$ is the maximum response, $\lambda_2$ is the dose with 50\% of the maximal response, $m$ is the effective concentration, and $\lambda_3$ is the effect of amlodipine not through the effective concentration (e.g., changes in behavior due to side-effects). 

Now I show that this model is vacuous. First, $f$ is allowed to be any probability density function that satisfies the constraint of the context. Next, if both $\theta_0$ and $\theta_1$ are non-negative, then $g(\theta)$ spans any non-negative drug concentration. Finally, note the image of $h$ is $[0,1]$ for any $b,a,m$ when each parameter in $\lambda$ is allowed to lie within $(-\infty, \infty)$ besides $\lambda_2 \ge 0$. Having shown the model to be vacuous, attention is turned to imposing assumptions on the parameters.

First, a choice for $f$ is made. Here, I use the first measured systolic blood pressure from the National Health and Nutrition Survey (NHANES) 2017-2018 \cite{NHANESQuestionnairesDatasets}. Rather than rely on a parametric simplification, the empirical distribution of these values is used. Next, I will assume that the doses are correctly measured such that $\theta_0 := 10$ when $A=1$ and $\theta_0 := 0$ when $A=0$. Further, I will assume that the drug concentration does not increase over the 24-hour period, $\theta_1 \in [0,1]$, so that the effective concentration is in the interval $[0, \theta_0]$. To further reduce the range of $\theta_1$, I use reported pharmacokinetic data on amlodipine \cite{kimPharmacokineticComparison22013, heoQuantitativeModelBlood2016}. Informed by the interval of concentrations at 24 hours in Figure 1 of \cite{kimPharmacokineticComparison22013}, a choice of $\Theta_1^* := [0.25, 0.40]$ was made. For the hypertension model, $\lambda_0$ is assumed to be zero, or that there is no other change in systolic blood pressure except through $A$. Additionally, I assume that amlodipine has no effect on hypertension except through the effective concentration at 24 hours, i.e., $\lambda_3 := 0$. Previously reported pharmacodynamic results informed the choices of $\Lambda_1^* := [16.3, 36.3]$ \cite{parkPharmacokineticHaemodynamicInteractions2019}, and $\Lambda_2^* := [0.1, 13.0]$ \cite{heoQuantitativeModelBlood2016}. To approximate the integral in \eqref{Eq3} with the given parameter values, each observation from NHANES has hypertension predicted from the effective concentration and the above Hill Equation model. The sample-weighted average of the predicted hypertension is then taken. To compute the bounds $\Psi^*$, the minimum and maximum values were found by computing $\bar{\mu}_a$ for each unique parameter combination. The estimated bounds were $\Psi^* =[0.23, 0.91]$, which indicates a substantial protective effect of amlodipine on hypertension in the population.

\subsection*{Model Evaluation}

Having constructed and applied the model, now consider how the proposed framework can be used to judge this model and make recommendations for its improvement. A first concern revolves around the context. The scientific question addresses the place (United States) but not the time. Greater specificity of the scientific question is needed to assess whether the NHANES 2017-2018 data to parameterize $f(w)$ is reasonable. A second challenge is that the NHANES data only provides an \textit{estimate} of the probability density function (albeit nonparametric). Within the mathematical model, the statistical uncertainty of NHANES being a sample is not incorporated into this mathematical model.

For $g$, it is assumed that the dose given was exactly as prescribed. However, there may be manufacturing uncertainty or non-compliance. The previous model could easily incorporate either by having $\theta_0$ be specified as a range instead. For choices of $\theta_1$, this information came from a study on male individuals from South Korea with a systolic blood pressure below 150. However, this differs from the context $S=1$ (e.g., United States, systolic blood pressure above 140, not restricted by gender). A potentially important difference between contexts is body weight distribution since body weight is important for effective concentrations of amlodipine \cite{heoQuantitativeModelBlood2016}. Specifically, the mean body weight was approximately 68 kg in the South Korean data while it was 84 kg in the NHANES data. To address this difference between contexts, either the model for the effective concentration function could be extended to incorporate body weight, or one could use generalizability methods from statistical modeling to generalize the effective concentration data to the body weight distribution of the United States data \cite{westreichTransportabilityTrialResults2017}.

For $h$, the choice $\lambda_0 := 0$ seems reasonable but that choice means that $\bar{\mu}_0 = 0$ regardless of all other parameters. Given that systolic blood pressures have some fluctuation, a range of values centered around zero for $\lambda_0$ may be better motivated. Like the effective concentration model, the ranges for $\lambda_1$ and $\lambda_2$ were based on data from individuals from South Korea with a systolic blood pressure less than 150. If the effectiveness of amlodipine, as defined by these parameters, varies substantially by baseline systolic blood pressure or body weight, then these ranges are suspect. Finally, the choice of $\lambda_3 := 0$ also may seem reasonable, but side-effects of amlodipine (e.g., nausea, fatigue, dizziness) may alter other behaviors. Therefore, exploring \edit{deviations from} zero could be beneficial. Finally, note that $\bar{y}$ can return negative values but systolic blood pressure is non-negative. Choosing another function for $\bar{y}$ that has a non-negative image may be preferred, otherwise care needs to be applied to ensure that the choices of $\lambda$ ensure non-negative outputs.

Altogether, this example illustrates how the identification framework can be used to evaluate models. Here, several issues with the pharmacodynamic model arose in specification of the scientific question, choice of the model structures, and the range of parameter values. the proposed framework also suggest how each issue could be remedied.

\section*{Conclusions}

Causal inference provides a set of tools for understanding the relation between mathematical manipulations performed on a sequence of numbers and sentences that express causal conclusions \cite{robinsAssociationCausationMarginal1999}. The utility of causal inference has been recognized in the statistical modeling tradition. Here, I have shown how concepts from causal inference might be used to analyze and clarify the assumptions underlying applications of mathematical models in biological, physical, and social sciences. \edit{This perspective illustrates how both statistical and mathematical modeling can be viewed as ways to bound causal effects for a context. The width of these bounds depends on the strength of the assumptions made. A shared framework between these traditions provides a basis for comparison when they are separately used to address a common scientific question. Assumptions of each modeling strategy can also be placed on equal footing. Finally, a shared framework basis can help to clarify how statistical and mathematical modeling can be jointly used to address a common question, something already done in practice \cite{zivichSynthesisEstimatorsTransportability2025, rahmstorfSemiEmpiricalApproachProjecting2007, wrightSemiEmpiricalModellingSLD, sausenEfficiencyMaximizationFixedbed2018, rezaeiHybridPhenomenologicalMathematicalBased2020}.}

As this work is an early attempt at a formal framework for theoretical analysis of mathematical models from a causal inference perspective, there are numerous questions remaining and extensions to be made. Here, I highlight some of these areas. First, my approach only produces bounds for $\psi$ with a given mathematical model. However, it should be relatively easy to replace the mathematical model parameter vector with an ordered pair, where the first element is the value for the parameter and the second element is the information or `belief' that supports that parameter value choice. With this extension, the mathematical model can then produce distributions for, rather than bounds on, $\psi$. Second, the models here assumed units were independent, but mathematical models can easily incorporate interactions between units \cite{edwardsInvitedCommentaryCausal2017b}. To adapt the approach used here, one would need to use an extension of potential outcomes for interference \cite{hudgensCausalInferenceInterference2008a}. \edit{After that, one would need to specify} how connections \edit{or interactions} between units occur in the \edit{desired context}. Similarly, this framework does not currently handle differential equations. Differential equations play an important role in the mathematical modeling tradition \cite{aalenCanWeBelieve2016}. \edit{An extension to differential equations could draw inspiration from work on continuous-time processes in the causal inference literature \cite{gillCausalInferenceComplex2001, zhangCAUSALINFERENCECONTINUOUSTIME2011, sunCausalIdentificationContinuoustime2022b}. Alternatively, one could approach this problem by imposing a discrete time-order, as done with the Euler method for solving ordinary differential equations}. Third, it is common within statistical modeling to compare predictions under the `natural course', or setting $A$ as it occurs in the world \cite{rudolphRoleNaturalCourse2021}, and other observed data implications as a check for gross violations of the underlying identification and estimation assumptions \cite{dahabrehCausallyInterpretableMetaanalysis2020a, keilParametricGformulaTimetoevent2014, zivichBridgedTreatmentComparisons2025, shook-saFusingTrialData2024}. In the mathematical modeling literature, comparing model output to observations is often referred to as calibration \cite{mccullochCalibratingAgentBasedModels2022, ogaraImprovingPolicyorientedAgentbased2025}. The notion of identification presented here may help to better understand and formalize calibration procedures for mathematical models. Finally, the motivating examples were in the context of medical science applications. Applying causal inference methods and this framework for quantitative analyses in other scientific fields is of interest.

\section*{Acknowledgements}

Financial Support: The author of this work was supported by grants from the National Institute of Allergy and Infectious Diseases (K01AI177102) and the National Institute of General Medical Sciences (R01GM140564). The content is solely the responsibility of the author and does not necessarily represent the official views of the National Institutes of Health.
~\\~\\
\noindent
Data and Code: Data and code used for the illustrative examples and the simulation experiment are publicly available on GitHub at \url{https://github.com/pzivich/publications-code}.
~\\~\\
\noindent
Thanks to Joseph C Lemaitre, Stephen R Cole, Jessie K Edwards, Eric T Lofgren, University of North Carolina at Chapel Hill Infectious Disease Dynamics group for their feedback on earlier iterations of these ideas and manuscript. Acknowledging them here is to express my gratitude and does not imply their endorsement.

\small
\bibliography{MathModels2}{}

\begin{thebibliography}{10}

\bibitem{mayUsesAbusesMathematics2004}
R.~M. May, ``Uses and {{Abuses}} of {{Mathematics}} in {{Biology}},'' {\em
  Science}, vol.~303, pp.~790--793, Feb. 2004.

\bibitem{kuhneCausalEvidenceHealth2022}
F.~K{\"u}hne, M.~Schomaker, I.~Stojkov, B.~Jahn, A.~{Conrads-Frank},
  S.~Siebert, G.~Sroczynski, S.~Puntscher, D.~Schmid, P.~{Schnell-Inderst}, and
  U.~Siebert, ``Causal evidence in health decision making: Methodological
  approaches of causal inference and health decision science,'' {\em German
  medical science: GMS e-journal}, vol.~20, p.~Doc12, 2022.

\bibitem{pengComparisonPhenomenologicalApproach2022}
Q.~Peng, W.~S. Gorter, and F.~J. Vermolen, ``Comparison between a
  phenomenological approach and a morphoelasticity approach regarding the
  displacement of extracellular matrix,'' {\em Biomechanics and Modeling in
  Mechanobiology}, vol.~21, no.~3, pp.~919--935, 2022.

\bibitem{naimiCommentaryIntegratingComplex2016a}
A.~I. Naimi, ``Commentary: {{Integrating Complex Systems Thinking}} into
  {{Epidemiologic Research}},'' {\em Epidemiology}, vol.~27, pp.~843--847, Nov.
  2016.

\bibitem{transtrumBridgingMechanisticPhenomenological2016}
M.~K. Transtrum and P.~Qiu, ``Bridging {{Mechanistic}} and {{Phenomenological
  Models}} of {{Complex Biological Systems}},'' {\em PLoS Computational
  Biology}, vol.~12, p.~e1004915, May 2016.

\bibitem{weinbergerMakingSenseNonfactual2020}
N.~Weinberger and S.~Bradley, ``Making sense of non-factual disagreement in
  science,'' {\em Studies in History and Philosophy of Science Part A},
  vol.~83, pp.~36--43, Oct. 2020.

\bibitem{whiteShouldWeCare2019}
C.~R. White and D.~J. Marshall, ``Should {{We Care If Models Are
  Phenomenological}} or {{Mechanistic}}?,'' {\em Trends in Ecology \&
  Evolution}, vol.~34, pp.~276--278, Apr. 2019.

\bibitem{christopheComputationalModelsNeurosciences2022}
G.~Christophe, C.~Brun, T.~Boraud, M.~Carlu, and D.~Depannemaecker,
  ``Computational models in neurosciences between mechanistic and
  phenomenological characterizations.'' Jan. 2022.

\bibitem{bakerMechanisticModelsMachine2018}
R.~E. Baker, J.-M. Pe{\~n}a, J.~Jayamohan, and A.~J{\'e}rusalem, ``Mechanistic
  models versus machine learning, a fight worth fighting for the biological
  community?,'' {\em Biology Letters}, vol.~14, p.~20170660, May 2018.

\bibitem{naimiIntroductionMethods2017}
A.~I. Naimi, S.~R. Cole, and E.~H. Kennedy, ``An introduction to g methods,''
  {\em International journal of epidemiology}, vol.~46, no.~2, pp.~756--762,
  2017.

\bibitem{robinsNewApproachCausal1986}
J.~Robins, ``A new approach to causal inference in mortality studies with a
  sustained exposure period---application to control of the healthy worker
  survivor effect,'' {\em Mathematical modelling}, vol.~7, no.~9-12,
  pp.~1393--1512, 1986.

\bibitem{bjornstadModelingInfectiousEpidemics2020}
O.~N. Bj{\o}rnstad, K.~Shea, M.~Krzywinski, and N.~Altman, ``Modeling
  infectious epidemics,'' {\em Nature Methods}, vol.~17, pp.~455--456, May
  2020.

\bibitem{bjornstadSEIRSModelInfectious2020}
O.~N. Bj{\o}rnstad, K.~Shea, M.~Krzywinski, and N.~Altman, ``The {{SEIRS}}
  model for infectious disease dynamics,'' {\em Nature Methods}, vol.~17,
  pp.~557--558, June 2020.

\bibitem{krijkampMicrosimulationModelingHealth2018}
E.~M. Krijkamp, F.~{Alarid-Escudero}, E.~A. Enns, H.~J. Jalal, M.~M. Hunink,
  and P.~Pechlivanoglou, ``Microsimulation modeling for health decision
  sciences using {{R}}: A tutorial,'' {\em Medical decision making : an
  international journal of the Society for Medical Decision Making}, vol.~38,
  pp.~400--422, Apr. 2018.

\bibitem{caglayanMicrosimulationModelingOncology2018}
{\c C}.~{\c C}a{\u g}layan, H.~Terawaki, Q.~Chen, A.~Rai, T.~Ayer, and C.~R.
  Flowers, ``Microsimulation {{Modeling}} in {{Oncology}},'' {\em JCO Clinical
  Cancer Informatics}, vol.~2, p.~CCI.17.00029, Mar. 2018.

\bibitem{el-sayedSocialNetworkAnalysis2012b}
A.~M. {El-Sayed}, P.~Scarborough, L.~Seemann, and S.~Galea, ``Social network
  analysis and agent-based modeling in social epidemiology,'' {\em
  Epidemiologic Perspectives \& Innovations}, vol.~9, p.~1, Feb. 2012.

\bibitem{tedeschiReviewPrevailingMathematical2023}
L.~O. Tedeschi, ``Review: {{The}} prevailing mathematical modeling
  classifications and paradigms to support the advancement of sustainable
  animal production,'' {\em animal}, vol.~17, p.~100813, Dec. 2023.

\bibitem{tolasaTheoreticalAnalysisSimple2025}
D.~G. Tolasa, ``Theoretical {{Analysis}} of a {{Simple Pendulum Experiment}},''
  {\em International Journal of Current Research in Science, Engineering \&
  Technology}, vol.~8, pp.~214--218, Apr. 2025.

\bibitem{bauerQuietRevolutionNumerical2015}
P.~Bauer, A.~Thorpe, and G.~Brunet, ``The quiet revolution of numerical weather
  prediction,'' {\em Nature}, vol.~525, pp.~47--55, Sept. 2015.

\bibitem{zhangMachineLearningMethods2025}
H.~Zhang, Y.~Liu, C.~Zhang, and N.~Li, ``Machine {{Learning Methods}} for
  {{Weather Forecasting}}: {{A Survey}},'' {\em Atmosphere}, vol.~16, p.~82,
  Jan. 2025.

\bibitem{heesterbeekModelingInfectiousDisease2015}
H.~Heesterbeek, R.~M. Anderson, V.~Andreasen, S.~Bansal, D.~De~Angelis, C.~Dye,
  K.~T.~D. Eames, W.~J. Edmunds, S.~D.~W. Frost, S.~Funk, T.~D. Hollingsworth,
  T.~House, V.~Isham, P.~Klepac, J.~Lessler, J.~O. {Lloyd-Smith}, C.~J.~E.
  Metcalf, D.~Mollison, L.~Pellis, J.~R.~C. Pulliam, M.~G. Roberts, C.~Viboud,
  and {ISAAC NEWTON INSTITUTE IDD COLLABORATION}, ``Modeling infectious disease
  dynamics in the complex landscape of global health,'' {\em Science},
  vol.~347, p.~aaa4339, Mar. 2015.

\bibitem{miyamaPhenomenologicalMechanisticModels2022}
T.~Miyama, S.-M. Jung, K.~Hayashi, A.~Anzai, R.~Kinoshita, T.~Kobayashi, N.~M.
  Linton, A.~Suzuki, Y.~Yang, B.~Yuan, T.~Kayano, A.~R. Akhmetzhanov, and
  H.~Nishiura, ``Phenomenological and mechanistic models for predicting early
  transmission data of {{COVID-19}},'' {\em Mathematical biosciences and
  engineering: MBE}, vol.~19, pp.~2043--2055, Jan. 2022.

\bibitem{faganStemCellsSystems2016}
M.~B. Fagan, ``Stem cells and systems models: Clashing views of explanation,''
  {\em Synthese}, vol.~193, pp.~873--907, Mar. 2016.

\bibitem{arbogastImpactComprehensiveWorkplace2016a}
J.~W. Arbogast, L.~{Moore-Schiltz}, W.~R. Jarvis, A.~{Harpster-Hagen},
  J.~Hughes, and A.~Parker, ``Impact of a {{Comprehensive Workplace Hand
  Hygiene Program}} on {{Employer Health Care Insurance Claims}} and {{Costs}},
  {{Absenteeism}}, and {{Employee Perceptions}} and {{Practices}},'' {\em
  Journal of Occupational and Environmental Medicine}, vol.~58, pp.~e231--e240,
  June 2016.

\bibitem{beamerModelingHumanViruses2015a}
P.~I. Beamer, K.~R. Plotkin, C.~P. Gerba, L.~Y. Sifuentes, D.~W. Koenig, and
  K.~A. Reynolds, ``Modeling of human viruses on hands and risk of infection in
  an office workplace using micro-activity data,'' {\em J Occup Environ Hyg},
  vol.~12, no.~4, pp.~266--75, 2015.

\bibitem{holfordClinicalTrialSimulation2010b}
N.~Holford, S.~C. Ma, and B.~A. Ploeger, ``Clinical {{Trial Simulation}}: {{A
  Review}},'' {\em Clinical Pharmacology \& Therapeutics}, vol.~88, no.~2,
  pp.~166--182, 2010.

\bibitem{ackleyDynamicalModelingTool2022}
S.~F. Ackley, J.~Lessler, and M.~M. Glymour, ``Dynamical {{Modeling}} as a
  {{Tool}} for {{Inferring Causation}},'' {\em American Journal of
  Epidemiology}, vol.~191, pp.~1--6, Jan. 2022.

\bibitem{breimanStatisticalModelingTwo2001a}
L.~Breiman, ``Statistical {{Modeling}}: {{The Two Cultures}},'' {\em
  Statistical Science}, vol.~16, pp.~199--231, Aug. 2001.

\bibitem{kennedyCommentStatisticalModeling2021}
E.~H. Kennedy, M.~Bonvini, and A.~Mishler, ``Comment on "{{Statistical
  Modeling}}: {{The Two Cultures}}" by {{Leo Breiman}},'' {\em Observational
  Studies}, vol.~7, no.~1, pp.~145--156, 2021.

\bibitem{murrayComparisonAgentBasedModels2017}
E.~J. Murray, J.~M. Robins, G.~R. Seage, K.~A. Freedberg, and M.~A. Hernan, ``A
  {{Comparison}} of {{Agent-Based Models}} and the {{Parametric G-Formula}} for
  {{Causal Inference}},'' {\em Am J Epidemiol}, vol.~186, pp.~131--142, July
  2017.

\bibitem{edwardsInvitedCommentaryCausal2017}
J.~K. Edwards, C.~R. Lesko, and A.~P. Keil, ``Invited {{Commentary}}: {{Causal
  Inference Across Space}} and {{Time-Quixotic Quest}}, {{Worthy Goal}}, or
  {{Both}}?,'' {\em Am J Epidemiol}, vol.~186, pp.~143--145, July 2017.

\bibitem{keyesInvitedCommentaryAgentBased2017c}
K.~M. Keyes, M.~Tracy, S.~J. Mooney, A.~Shev, and M.~Cerd{\'a}, ``Invited
  {{Commentary}}: {{Agent-Based Models-Bias}} in the {{Face}} of
  {{Discovery}},'' {\em American Journal of Epidemiology}, vol.~186,
  pp.~146--148, July 2017.

\bibitem{mooneyGComputationAgentBasedModeling2022}
S.~J. Mooney, A.~B. Shev, K.~M. Keyes, M.~Tracy, and M.~Cerd{\'a},
  ``G-{{Computation}} and {{Agent-Based Modeling}} for {{Social Epidemiology}}:
  {{Can Population Interventions Prevent Posttraumatic Stress Disorder}}?,''
  {\em American Journal of Epidemiology}, vol.~191, pp.~188--197, Jan. 2022.

\bibitem{arnoldDAGinformedRegressionModelling2018}
K.~F. Arnold, W.~J. Harrison, A.~J. Heppenstall, and M.~S. Gilthorpe,
  ``{{DAG-informed}} regression modelling, agent-based modelling and
  microsimulation modelling: A critical comparison of methods for causal
  inference,'' {\em Int J Epidemiol}, Dec. 2018.

\bibitem{mooijOrdinaryDifferentialEquations2013}
J.~M. Mooij, D.~Janzing, and B.~Sch{\"o}lkopf, ``From {{Ordinary Differential
  Equations}} to {{Structural Causal Models}}: The deterministic case,'' Apr.
  2013.

\bibitem{aalenCanWeBelieve2016}
{\relax OO}.~Aalen, K.~R{\o}ysland, {\relax JM}.~Gran, R.~Kouyos, and T.~Lange,
  ``Can we believe the {{DAGs}}? {{A}} comment on the relationship between
  causal {{DAGs}} and mechanisms,'' {\em Statistical Methods in Medical
  Research}, vol.~25, pp.~2294--2314, Oct. 2016.

\bibitem{wordenProductsCompartmentalModels2017}
L.~Worden and T.~C. Porco, ``Products of {{Compartmental Models}} in
  {{Epidemiology}},'' {\em Computational and Mathematical Methods in Medicine},
  vol.~2017, p.~8613878, 2017.

\bibitem{havumakiUsingCompartmentalModels2020}
J.~Havumaki and M.~C. Eisenberg, ``Using compartmental models to simulate
  directed acyclic graphs to explore competing causal mechanisms underlying
  epidemiological study data,'' {\em Journal of The Royal Society Interface},
  vol.~17, p.~20190675, June 2020.

\bibitem{marshallFormalizingRoleAgentBased2015}
B.~D.~L. Marshall and S.~Galea, ``Formalizing the {{Role}} of {{Agent-Based
  Modeling}} in {{Causal Inference}} and {{Epidemiology}},'' {\em American
  Journal of Epidemiology}, vol.~181, pp.~92--99, Jan. 2015.

\bibitem{zivichSynthesisEstimatorsTransportability2025}
P.~N. Zivich, J.~K. Edwards, B.~E. {Shook-Sa}, E.~T. Lofgren, J.~Lessler, and
  S.~R. Cole, ``Synthesis estimators for transportability with positivity
  violations by a continuous covariate,'' {\em Journal of the Royal Statistical
  Society Series A: Statistics in Society}, vol.~188, pp.~158--180, Jan. 2025.

\bibitem{dahabrehCausalInferenceEffects2024}
I.~J. Dahabreh and K.~{Bibbins-Domingo}, ``Causal {{Inference About}} the
  {{Effects}} of {{Interventions From Observational Studies}} in {{Medical
  Journals}},'' {\em JAMA}, vol.~331, pp.~1845--1853, June 2024.

\bibitem{imbensCausalInferenceSocial2024}
G.~W. Imbens, ``Causal {{Inference}} in the {{Social Sciences}},'' {\em Annual
  Review of Statistics and Its Application}, vol.~11, pp.~123--152, Apr. 2024.

\bibitem{itoDistinguishingDescriptionPrediction2025}
C.~Ito, L.~{Al-Hassany}, T.~Kurth, and T.~Glatz, ``Distinguishing
  {{Description}}, {{Prediction}}, and {{Causal Inference}},'' {\em Neurology},
  vol.~104, p.~e210171, Feb. 2025.

\bibitem{wintherMathematicalModelingBiology2012}
R.~G. Winther, ``Mathematical {{Modeling}} in {{Biology}}: {{Philosophy}} and
  {{Pragmatics}},'' {\em Frontiers in Plant Science}, vol.~3, June 2012.

\bibitem{manskiNonparametricBoundsTreatment1990}
C.~F. Manski, ``Nonparametric {{Bounds}} on {{Treatment Effects}},'' {\em The
  American Economic Review}, vol.~80, no.~2, pp.~319--323, 1990.

\bibitem{dangCausalRoadmapGenerating2023}
L.~E. Dang, S.~Gruber, H.~Lee, I.~J. Dahabreh, E.~A. Stuart, B.~D. Williamson,
  R.~Wyss, I.~D{\'i}az, D.~Ghosh, E.~K{\i}c{\i}man, D.~Alemayehu, K.~L.
  Hoffman, C.~Y. Vossen, R.~A. Huml, H.~Ravn, K.~Kvist, R.~Pratley, M.-C. Shih,
  G.~Pennello, D.~Martin, S.~P. Waddy, C.~E. Barr, M.~Akacha, J.~B. Buse,
  M.~van~der Laan, and M.~Petersen, ``A causal roadmap for generating
  high-quality real-world evidence,'' {\em Journal of Clinical and
  Translational Science}, vol.~7, p.~e212, Jan. 2023.

\bibitem{hudgensCausalInferenceInterference2008a}
M.~G. Hudgens and M.~E. Halloran, ``Toward {{Causal Inference With
  Interference}},'' {\em Journal of the American Statistical Association},
  vol.~103, no.~482, pp.~832--842, 2008.

\bibitem{savjeAverageTreatmentEffects2021}
F.~S{\"a}vje, P.~M. Aronow, and M.~G. Hudgens, ``Average treatment effects in
  the presence of unknown interference,'' {\em The Annals of Statistics},
  vol.~49, pp.~673--701, Apr. 2021.

\bibitem{splawa-neymanApplicationProbabilityTheory1990}
J.~{Splawa-Neyman}, D.~M. Dabrowska, and {\relax TP}.~Speed, ``On the
  application of probability theory to agricultural experiments. {{Essay}} on
  principles. {{Section}} 9,'' {\em Statistical Science}, pp.~465--472, 1990.

\bibitem{palAnimalHumanTetanus2024}
M.~Pal, T.~Rebuma, M.~Regassa, and F.~Tariku, ``Animal and human tetanus:
  {{An}} overview on transmission, pathogenesis, epidemiology, diagnosis, and
  control,'' {\em Journal of Advances in Microbiological Research}, vol.~5,
  no.~1, pp.~22--26, 2024.

\bibitem{kolmogorovFoundationsTheoryProbability1950}
A.~N. Kolmogorov, {\em Foundations of the Theory of Probability}.
\newblock New York: Chelsea Pub. Co., 1950.

\bibitem{aronowNonparametricIdentificationNot2025}
P.~M. Aronow, J.~M. Robins, T.~Saarinen, F.~S{\"a}vje, and J.~Sekhon,
  ``Nonparametric identification is not enough, but randomized controlled
  trials are,'' {\em Observational Studies}, vol.~11, no.~1, pp.~3--16, 2025.

\bibitem{robinsAssociationCausationMarginal1999}
J.~M. Robins, ``Association, {{Causation}}, and {{Marginal Structural
  Models}},'' {\em Synthese}, vol.~121, no.~1/2, pp.~151--179, 1999.

\bibitem{eisenbergDeterminingIdentifiableParameter2014}
M.~C. Eisenberg and M.~A.~L. Hayashi, ``Determining identifiable parameter
  combinations using subset profiling,'' {\em Mathematical Biosciences},
  vol.~256, pp.~116--126, Oct. 2014.

\bibitem{kabanikhinIdentifiabilityMathematicalModels2016}
S.~I. Kabanikhin, D.~A. Voronov, A.~A. Grodz, and O.~I. Krivorotko,
  ``Identifiability of mathematical models in medical biology,'' {\em Russian
  Journal of Genetics: Applied Research}, vol.~6, pp.~838--844, Dec. 2016.

\bibitem{krivorotkoSensitivityAnalysisPractical2020}
O.~I. Krivorotko, D.~V. Andornaya, and S.~I. Kabanikhin, ``Sensitivity
  {{Analysis}} and {{Practical Identifiability}} of {{Some Mathematical
  Models}} in {{Biology}},'' {\em Journal of Applied and Industrial
  Mathematics}, vol.~14, pp.~115--130, Jan. 2020.

\bibitem{kabanikhinPracticalIdentifiabilityMathematical2021}
S.~Kabanikhin, M.~Bektemesov, O.~Krivorotko, and Z.~Bektemessov, ``Practical
  identifiability of mathematical models of biomedical processes,'' {\em
  Journal of Physics: Conference Series}, vol.~2092, p.~012014, Dec. 2021.

\bibitem{carpioParameterIdentificationEpidemiological2022}
A.~Carpio and E.~Pierret, ``Parameter identification in epidemiological
  models,'' {\em Mathematical Analysis of Infectious Diseases}, pp.~103--124,
  2022.

\bibitem{coleConsistencyStatementCausal2009}
S.~R. Cole and C.~E. Frangakis, ``The consistency statement in causal
  inference: A definition or an assumption?,'' {\em Epidemiology}, vol.~20,
  no.~1, pp.~3--5, 2009.

\bibitem{vanderweeleConcerningConsistencyAssumption2009}
T.~J. VanderWeele, ``Concerning the consistency assumption in causal
  inference,'' {\em Epidemiology (Cambridge, Mass.)}, vol.~20, pp.~880--883,
  Nov. 2009.

\bibitem{hernanEstimatingCausalEffects2006}
M.~A. Hern{\'a}n and J.~M. Robins, ``Estimating causal effects from
  epidemiological data,'' {\em Journal of Epidemiology \& Community Health},
  vol.~60, pp.~578--586, July 2006.

\bibitem{tchetgenIntroductionProximalCausal2024a}
E.~J.~T. Tchetgen, A.~Ying, Y.~Cui, X.~Shi, and W.~Miao, ``An {{Introduction}}
  to {{Proximal Causal Inference}},'' {\em Statistical Science}, vol.~39,
  pp.~375--390, Aug. 2024.

\bibitem{zivichINTRODUCINGPROXIMALCAUSAL2023}
P.~N. Zivich, S.~R. Cole, J.~K. Edwards, G.~E. Mulholland, B.~E. {Shook-Sa},
  and E.~J. Tchetgen~Tchetgen, ``{{INTRODUCING PROXIMAL CAUSAL INFERENCE FOR
  EPIDEMIOLOGISTS}},'' {\em American Journal of Epidemiology}, vol.~192,
  pp.~1224--1227, July 2023.

\bibitem{vansteelandtModelSelectionModel2012}
S.~Vansteelandt, M.~Bekaert, and G.~Claeskens, ``On model selection and model
  misspecification in causal inference,'' {\em Statistical Methods in Medical
  Research}, vol.~21, pp.~7--30, Feb. 2012.

\bibitem{maclarenWhatCanBe2020}
O.~J. Maclaren and R.~Nicholson, ``What can be estimated? {{Identifiability}},
  estimability, causal inference and ill-posed inverse problems,'' {\em
  arXiv:1904.02826 [cs, math, stat]}, July 2020.

\bibitem{bangDoublyRobustEstimation2005a}
H.~Bang and J.~M. Robins, ``Doubly robust estimation in missing data and causal
  inference models,'' {\em Biometrics}, vol.~61, no.~4, pp.~962--973, 2005.

\bibitem{zivichMachineLearningCausal2022}
P.~N. Zivich, A.~Breskin, and E.~H. Kennedy, ``Machine {{Learning}} and
  {{Causal Inference}},'' in {\em Wiley {{StatsRef}}: {{Statistics Reference
  Online}}}, pp.~1--8, John Wiley \& Sons, Ltd, 2022.

\bibitem{westreichBerksonsBiasSelection2012}
D.~Westreich, ``Berkson's bias, selection bias, and missing data,'' {\em
  Epidemiology (Cambridge, Mass.)}, vol.~23, no.~1, pp.~159--164, 2012.

\bibitem{edwardsAllYourData2015}
J.~K. Edwards, S.~R. Cole, and D.~Westreich, ``All your data are always
  missing: Incorporating bias due to measurement error into the potential
  outcomes framework,'' {\em International Journal of Epidemiology}, vol.~44,
  pp.~1452--1459, Aug. 2015.

\bibitem{robertsConceptualizingModelReport2012}
M.~Roberts, L.~B. Russell, A.~D. Paltiel, M.~Chambers, P.~McEwan, and M.~Krahn,
  ``Conceptualizing a {{Model}}: {{A Report}} of the {{ISPOR-SMDM Modeling Good
  Research Practices Task Force}}--2,'' {\em Medical Decision Making}, vol.~32,
  pp.~678--689, Sept. 2012.

\bibitem{robinsonTenRulesEffective2022}
P.~A. Robinson, ``Ten rules for effective modeling,'' {\em NeuroImage},
  vol.~263, p.~119622, Nov. 2022.

\bibitem{padmanabhanModelingHowAntibody2022}
P.~Padmanabhan, R.~Desikan, and N.~M. Dixit, ``Modeling how antibody responses
  may determine the efficacy of {{COVID-19}} vaccines,'' {\em Nature
  Computational Science}, vol.~2, pp.~123--131, Feb. 2022.

\bibitem{mcgrathRevisitingGnullParadox2022}
S.~McGrath, J.~G. Young, and M.~A. Hern{\'a}n, ``Revisiting the g-null
  paradox,'' {\em Epidemiology (Cambridge, Mass.)}, vol.~33, pp.~114--120, Jan.
  2022.

\bibitem{coenHowMathematicalModels2007}
P.~G. Coen, ``How mathematical models have helped to improve understanding the
  epidemiology of infection,'' {\em Early Human Development}, vol.~83,
  pp.~141--148, Mar. 2007.

\bibitem{lofgrenReIntegratingComplex2017}
E.~T. Lofgren, B.~D. Marshall, and S.~Galea, ``Re: {{Integrating Complex
  Systems Thinking}} into {{Epidemiologic Research}},'' {\em Epidemiology},
  vol.~28, no.~5, p.~e50, 2017.

\bibitem{NHANESQuestionnairesDatasets}
``{{NHANES Questionnaires}}, {{Datasets}}, and {{Related Documentation}}.''
  https://wwwn.cdc.gov/nchs/nhanes/continuousnhanes/default.aspx?BeginYear=2017.

\bibitem{kimPharmacokineticComparison22013}
Y.~Kim, M.~Son, D.~Lee, H.~Roh, H.~Son, D.~Chae, M.~Y. Bahng, and K.~Park,
  ``Pharmacokinetic {{Comparison}} of 2 {{Fixed-Dose Combination Tablets}} of
  {{Amlodipine}} and {{Valsartan}} in {{Healthy Male Korean Volunteers}}: {{A
  Randomized}}, {{Open-Label}}, 2-{{Period}}, {{Single-Dose}}, {{Crossover
  Study}},'' {\em Clinical Therapeutics}, vol.~35, pp.~934--940, July 2013.

\bibitem{heoQuantitativeModelBlood2016}
Y.-A. Heo, N.~Holford, Y.~Kim, M.~Son, and K.~Park, ``Quantitative model for
  the blood pressure-lowering interaction of valsartan and amlodipine,'' {\em
  British Journal of Clinical Pharmacology}, vol.~82, no.~6, pp.~1557--1567,
  2016.

\bibitem{parkPharmacokineticHaemodynamicInteractions2019}
J.-W. Park, K.-A. Kim, Y.~Il~Kim, and J.-Y. Park, ``Pharmacokinetic and
  haemodynamic interactions between amlodipine and losartan in human beings,''
  {\em Basic \& Clinical Pharmacology \& Toxicology}, vol.~125, no.~4,
  pp.~345--352, 2019.

\bibitem{westreichTransportabilityTrialResults2017}
D.~Westreich, J.~K. Edwards, C.~R. Lesko, E.~Stuart, and S.~R. Cole,
  ``Transportability of {{Trial Results Using Inverse Odds}} of {{Sampling
  Weights}},'' {\em American Journal of Epidemiology}, vol.~186,
  pp.~1010--1014, Oct. 2017.

\bibitem{rahmstorfSemiEmpiricalApproachProjecting2007}
S.~Rahmstorf, ``A {{Semi-Empirical Approach}} to {{Projecting Future Sea-Level
  Rise}},'' {\em Science}, vol.~315, pp.~368--370, Jan. 2007.

\bibitem{wrightSemiEmpiricalModellingSLD}
W.~Wright and M.~Potapczuk, ``Semi-{{Empirical Modelling}} of {{SLD
  Physics}},'' in {\em 42nd {{AIAA Aerospace Sciences Meeting}} and
  {{Exhibit}}}, {American Institute of Aeronautics and Astronautics}.

\bibitem{sausenEfficiencyMaximizationFixedbed2018}
M.~G. Sausen, F.~B. Scheufele, H.~J. Alves, M.~G.~A. Vieira, M.~G.~C. {da
  Silva}, F.~H. Borba, and C.~E. Borba, ``Efficiency maximization of fixed-bed
  adsorption by applying hybrid statistical-phenomenological modeling,'' {\em
  Separation and Purification Technology}, vol.~207, pp.~477--488, Dec. 2018.

\bibitem{rezaeiHybridPhenomenologicalMathematicalBased2020}
R.~Rezaei, C.~Hayduk, E.~Alkan, T.~Kemski, T.~Delebinski, and C.~Bertram,
  ``Hybrid {{Phenomenological}} and {{Mathematical-Based Modeling Approach}}
  for {{Diesel Emission Prediction}},'' in {\em {{SAE Technical Paper}}},
  no.~2020-01-0660, Apr. 2020.

\bibitem{edwardsInvitedCommentaryCausal2017b}
J.~K. Edwards, C.~R. Lesko, and A.~P. Keil, ``Invited {{Commentary}}: {{Causal
  Inference Across Space}} and {{Time-Quixotic Quest}}, {{Worthy Goal}}, or
  {{Both}}?,'' {\em American Journal of Epidemiology}, vol.~186, pp.~143--145,
  July 2017.

\bibitem{gillCausalInferenceComplex2001}
R.~D. Gill and J.~M. Robins, ``Causal {{Inference}} for {{Complex Longitudinal
  Data}}: {{The Continuous Case}},'' {\em The Annals of Statistics}, vol.~29,
  no.~6, pp.~1785--1811, 2001.

\bibitem{zhangCAUSALINFERENCECONTINUOUSTIME2011}
M.~Zhang, M.~M. Joffe, and D.~S. Small, ``{{CAUSAL INFERENCE FOR
  CONTINUOUS-TIME PROCESSES WHEN COVARIATES ARE OBSERVED ONLY AT DISCRETE
  TIMES}},'' {\em Annals of statistics}, vol.~39, pp.~10.1214/10--AOS830, Feb.
  2011.

\bibitem{sunCausalIdentificationContinuoustime2022b}
J.~Sun and F.~W. Crawford, ``Causal identification for continuous-time
  stochastic processes,'' Nov. 2022.

\bibitem{rudolphRoleNaturalCourse2021}
J.~E. Rudolph, A.~Cartus, L.~M. Bodnar, E.~F. Schisterman, and A.~I. Naimi,
  ``The {{Role}} of the {{Natural Course}} in {{Causal Analysis}},'' {\em
  American Journal of Epidemiology}, p.~kwab248, Oct. 2021.

\bibitem{dahabrehCausallyInterpretableMetaanalysis2020a}
I.~J. Dahabreh, L.~C. Petito, S.~E. Robertson, M.~A. Hern{\'a}n, and J.~A.
  Steingrimsson, ``Toward {{Causally Interpretable Meta-analysis}}:
  {{Transporting Inferences}} from {{Multiple Randomized Trials}} to a {{New
  Target Population}},'' {\em Epidemiology (Cambridge, Mass.)}, vol.~31,
  pp.~334--344, May 2020.

\bibitem{keilParametricGformulaTimetoevent2014}
A.~P. Keil, J.~K. Edwards, D.~R. Richardson, A.~I. Naimi, and S.~R. Cole, ``The
  parametric {{G-formula}} for time-to-event data: Towards intuition with a
  worked example,'' {\em Epidemiology (Cambridge, Mass.)}, vol.~25, no.~6,
  pp.~889--897, 2014.

\bibitem{zivichBridgedTreatmentComparisons2025}
P.~N. Zivich, S.~R. Cole, J.~K. Edwards, B.~E. {Shook-Sa}, A.~Breskin, and
  M.~G. Hudgens, ``Bridged treatment comparisons: An illustrative application
  in {{HIV}} treatment,'' {\em American Journal of Epidemiology}, vol.~194,
  pp.~1687--1694, June 2025.

\bibitem{shook-saFusingTrialData2024}
B.~E. {Shook-Sa}, P.~N. Zivich, S.~P. Rosin, J.~K. Edwards, A.~A. Adimora,
  M.~G. Hudgens, and S.~R. Cole, ``Fusing trial data for treatment comparisons:
  {{Single}} vs multi-span bridging,'' {\em Statistics in Medicine}, vol.~43,
  no.~4, pp.~793--815, 2024.

\bibitem{mccullochCalibratingAgentBasedModels2022}
J.~McCulloch, J.~Ge, J.~A. Ward, A.~Heppenstall, J.~G. Polhill, and
  N.~Malleson, ``Calibrating {{Agent-Based Models Using Uncertainty
  Quantification Methods}},'' {\em Journal of Artificial Societies and Social
  Simulation}, vol.~25, no.~2, p.~1, 2022.

\bibitem{ogaraImprovingPolicyorientedAgentbased2025}
D.~O'Gara, C.~C. Kerr, D.~J. Klein, M.~Binois, R.~Garnett, and R.~A. Hammond,
  ``Improving policy-oriented agent-based modeling with history matching: {{A}}
  case study,'' {\em Epidemics}, vol.~52, p.~100845, Sept. 2025.

\end{thebibliography}
\bibliographystyle{ieeetr}

\newpage
\normalsize

\end{document}